\documentclass[aps,prl,preprint,groupedaddress]{revtex4-1}

\usepackage{amsmath}
\usepackage{amsfonts}
\usepackage{amssymb}
\usepackage{graphicx}
\usepackage{natbib}
\usepackage{color, soul}

\newcommand{\w}{\omega}

\begin{document}

\title{A practical scheme for generating isolated elliptically polarized attosecond pulses using bi-chromatic counter rotating circularly polarized laser fields}

\author{Lukas Medi\v sauskas}
\address{Department of Physics, Imperial College London, South Kensington Campus, SW7 2AZ London, United Kingdom}
\address{Max-Born-Institute, Max-Born Strasse 2A, D-12489 Berlin, Germany}

\author{Jack Wragg}
\address{Centre for Theoretical Atomic, Molecular and Optical Physics, School of Mathematics and Physics, Queenʼs University Belfast, Belfast BT7 1NN, UK}

\author{Hugo van der Hart}
\address{Centre for Theoretical Atomic, Molecular and Optical Physics, School of Mathematics and Physics, Queenʼs University Belfast, Belfast BT7 1NN, UK}

\author{Misha Yu. Ivanov}
\address{Max-Born-Institute, Max-Born Strasse 2A, D-12489 Berlin, Germany}
\address{Department of Physics, Humboldt University, Newtonstr. 15, D-12489 Berlin, Germany}
\address{Department of Physics, Imperial College London, South Kensington Campus, SW7 2AZ London, 
United Kingdom}

\date{\today}

\begin{abstract}
Spectra of circularly polarized harmonics is calculated by numerically solving the Time-Dependent Schr\" odinger Equation for a 2D model of Ne atom using circularly polarized fundamental with counter-rotating second harmonic laser fields. We demonstrate strong asymmetry between left- and right- circularly polarized harmonics when a ground state with p-type symmetry is used. It arises due to the circular polarization of individual attosecond pulses in the generated pulse train. Reducing the length of the counter-rotating drivers and introducing a small time-shift between them allows to generate a single elliptically polarized attosecond pulse.  
\end{abstract}

\maketitle

High Harmonic Generation (HHG) in atoms and molecules is a highly nonlinear process which 
up-converts intense infrared laser field into the extreme 
ultraviolet (XUV) and soft X-ray radiation \cite{Salières199983,Krausz2009,Corkum1993,Lewenstein1994}. 
The emitted light can be used to track quantum 
dynamics underlying the nonlinear response \cite{Haessler2010,Baker2006b,Worner2010,Smirnova2009}, or as a
table-top source of bright, coherent,
ultrashort pulses \cite{Cavalieri2007,Popmintchev2012,Salieres1999,Lepine2013}.

In the latter case,  generation 
of circular or highly elliptic high harmonics and/or attosecond XUV pulses  is 
very important.
Such pulses would find  numerous applications, e.g. in
chiral-sensitive light-matter interactions such as chiral recognition via photoelectron circular dichroism 
(PECD) (see e.g. \cite{Hergenhahn2004,Bowering2001,Powis2000}), 
study of ultrafast chiral-specific dynamics in molecules (e.g. \cite{Ferre2014,Travnikova2010}), and
X-ray Magnetic Circular Dichroism (XMCD) spectroscopy (e.g. \cite{Radu2011,Boeglin2010,Eisebitt2004,Fischer1999,Lopez-Flores2012,Stohr1993,Schutz1993}), including 
time-resolved imaging of magnetic structures (e.g. \cite{Radu2011,Boeglin2010,Eisebitt2004,Fischer1999,Lopez-Flores2012}).  
Table-top sources of sub-100 fs, or even attosecond, chiral pulses would be a real 
breakthrough for
laboratory-scale ultrafast studies.  Not surprisingly, search for schemes enabling the 
generation of short, coherent XUV pulses with tunable polarization is 
a very active area of research, see e.g. \cite{Chen2015,Morales2012,Yuan2013,Becker2000a,Becker2000,Yuan2011,Eichmann1995,Long1995,Ferre2014,Pisanty2014,Ivanov2014,Fleischer2014a,Kfir2014,Becker1999,Yuan2012a,Lambert2015}.

Importantly, the control over polarization is desired not only for individual harmonics, where it has just 
been demonstrated \cite{Kfir2014,Fleischer2014a}, but also
for  individual attosecond pulses, both isolated and in a train, where 
robust and practical scheme is still lacking. 
We show a way to solve this problem, proposing a practical scheme for
the generation of highly elliptic attosecond pulses, both single and in a train.

An elegant solution to generating individual high harmonics with circular polarization
has been found by W. Becker and  coworkers \cite{Eichmann1995,Long1995,Becker2000a,Becker2000}.
It relies on combining circularly polarized fundamental field with a counter-rotating second harmonic.
The resulting  electric field peaks three times within one cycle of the 
fundamental, producing three ionization bursts. 
The electron promoted to the continuum near the peak  of the instantaneous field 
can successfully revisit 
the parent ion within about half-cycle, emitting an attosecond radiation burst \cite{Becker2000a,Becker2000}.

This approach has now been very successfully used in \cite{Kfir2014,Fleischer2014a}, demonstrating 
generation of bright, phase matched high harmonic radiation. Importantly, 
tuning the ellipticity of one of the fields allows to tune the ellipticity of the generated high harmonics
from linear to circular \cite{Fleischer2014a}.
While the theoretical interpretation of this control is a matter of debate
\cite{Fleischer2014a,Pisanty2014}, the approach is very promising.
However, until now the possibility of 
extending this scheme from controlling the polarization of
individual harmonics to controlling the polarization of 
isolated attosecond pulses looked far from straightforward.

Indeed, the  driving field 
dictates that the direction of electron return rotates
by 120$^{0}$ three times per cycle. Consequently, recombination 
with an s-state yields three {\it linearly} polarized attosecond bursts per cycle, with
polarization rotating by 120$^{0}$ from burst to burst \cite{Becker2000}.
Thus, while each harmonic is circularly polarized,
the same does not apply to their superposition.

This can also be seen in the frequency domain. The harmonic lines are 
at energies $(n+1)\omega +2n\omega=(3n+1)
\omega$ and $n\omega+(n+1)2\omega=(3n+2)
\omega$. In centrally symmetric medium, and for circularly polarized
driving fields, the selection rules
dictate that the $\Omega=(3n+1)\omega$ line has the same circularity as 
the fundamental
while the $\Omega=(3n+2)\omega$ line has the same circularity as 
the second harmonic,
($\Omega=n\omega+2n\omega=3n\omega$ is parity forbidden)
\cite{Eichmann1995,Long1995,Alon1998,Fleischer2014a,Kfir2014,Mauger2014}.
Thus, the  harmonics have alternating helicity. Adding harmonics of alternating helicity
with equal intensity yields an attosecond pulse train where each subsequent pulse has 
linear polarization rotated by 120$^{0}$, in concert with 
the time domain picture. 

Suppressing every second allowed
harmonic line, e.g. $\Omega^{(3n+2)}=(3n+2)\omega $, would 
solve the problem of generating individual attosecond pulses with circular polarization.
O. Kfir et. al. \cite{Kfir2014} suggested that such suppression
can be achieved by
optimizing the phase-matching conditions in gas-filled hollow fiber and
reported substantial suppression of the lines $\Omega^{(3n+2)}=(3n+2)\omega $.

\begin{figure}[hb!]
    \centering
    \includegraphics[]{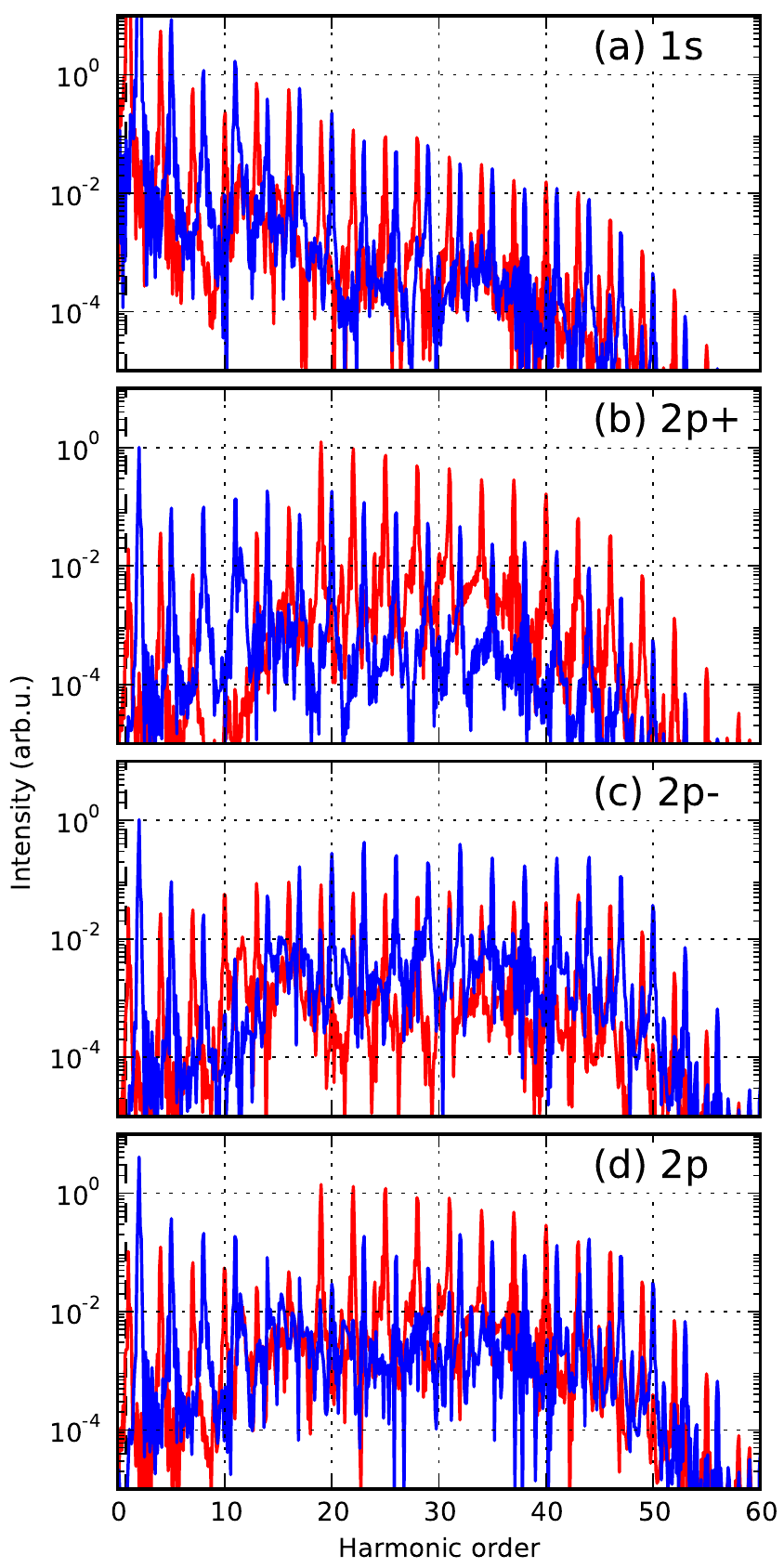}
    \caption{Spectra for (a) 1s, (b) 2p$_+$, (c) 2p$_-$ initial states and (d) equal 
mixture of 2p$_+$ and 2p$_-$ states. Colors mark harmonics 
co-rotating (red) and counter-rotating (blue) with the  $\w$ field. 
}
    \label{fig:spectra}
\end{figure}

First, we show that relative intensities of the counter-rotating harmonic lines
strongly depend on the orbital momentum of the initial state. 
For an initial p-state (as for a Neon , Argon, or Krypton gas), the harmonics 
co-rotating with the fundamental field can be much stronger then those co-rotating
with the second harmonic. The effect is found with the contribution of 
both degenerate sub-levels, $p+$ and $p-$, included in the calculation. 
As a result, circularly polarized
attosecond pulses are generated already at the microscopic, single-atom 
level, see Fig. \ref{fig:spectra}.
Additional help from phase-matching is a bonus, but not necessary.

Second, we extend the scheme to generation of isolated attosecond 
pulses. We show that when the counter-rotating driving pulses become relatively short,
e.g. 7-8 fs for the 800 nm driver and its second harmonic, one can generate an isolated attosecond pulse,
or a controllable train with 2 or 3 pulses, by tuning the time delay between the 
fundamental and the second harmonic.
 
To demonstrate these effects, we numerically solve the time dependent Schr\" oedinger equation (TDSE)
for a 2D Neon-like model atom, for counter-clockwise (+) polarized fundamental and 
clockwise (--) polarized second harmonic. 
We show that the harmonics generated from orbitals with 
m$=\pm1$ differ from those generated 
from  $s$ orbitals in two important ways. 
Firstly, the height of the adjacent left- and right- circularly polarized 
harmonics can differ by an order of magnitude, with m$=1$ state favouring 
harmonics co-rotating with fundamental and m$=-1$ state favouring harmonics co-rotating with $2\w$ field. 
Secondly, once the two contributions are added coherently, $+$ polarization
continues to dominate in a broad spectral range,  leading to highly elliptic circularly polarized 
attosecond pulse train already at the single-atom level.  
Our findings are in accord with \cite{Kfir2014}
(see Ne spectra in Fig. 3 of \cite{Kfir2014}), where
such disparity was attributed to  phase matching.

We solve the (TDSE) in the 
length gauge (atomic units are used throughout unless stated otherwise):
\begin{equation}
-\imath \frac{\partial}{\partial t}\Phi(t, r) = \left[\hat{T}+V(r)+\textbf{r} \cdot \textbf{E}(t)\right]\Phi(t, r).
\end{equation} 
The 2D model potential  is taken from \cite{Barth2014}
\begin{equation}
V(r) = -\frac{Z(r)}{\sqrt{r^2+a}} \label{eq:potential}
\end{equation}
where $Z(r)=1+9\exp(-r^2)$ and $a=2.88172$ to obtain the ionization potential of Ne atom $I_p=0.793$ a.u.
for the 2p orbitals. The 1s state has an energy $E_{1s}=-2.952$ a.u. and the 2s energy is $E_{2s}=-0.217$ a.u.. 
For reference calculations we use 1s as the initial state but keep 
the same ionization potential taking $Z(r)=1$ and $a=0.1195$.

The laser electric field is
\begin{align}
E(t) = &E_{ir}\cdot f(t)\cdot(\cos[\w t]+\cos[2\w t])\hat{x} \nonumber\\
+ &E_{ir}\cdot f(t)\cdot(\sin[\w t]-\sin[2\w t])\hat{y}
\end{align}
where $f(t)$ is the trapezoidal envelope with 2 cycle rising and falling edges and 5 cycle plateau (in units of fundamental).
The $\w$-field rotates counter-clockwise (+). The second harmonic rotates clockwise (--).

The TDSE is propagated on a 2D Cartesian grid using Taylor-series propagator 
with expansion up to 8th order \cite{Moler1978}. A complex absorbing potential 
\begin{equation}
V_{c}(x) = \eta \cdot (x-x_0)^n
\end{equation} 
with $\eta=5 \times 10^{-4}$ and $n=3$ is used to avoid non-physical reflections 
from the boundary. Other simulation parameters are summarized in Table \ref{table:parameters}. 

Convergence was tested with respect to the absorbing potential,
the time step and the spatial grid. Note that
HHG in bicircular fields is dominated by very short trajectories \cite{Becker2000a}.

\begin{table}
    \begin{tabular}{llll}
        \hline 
        laser frequency & $\w $ & 0.05 & ($\lambda = 911$ nm)\\
        laser electric field & $E_{ir}$ & 0.05 & (I$=0.88\cdot 10^{14}$ W/cm$^2$)\\
        & & \\
        grid step size & dr & 0.2 &\\
        time step size & dt & 0.005 &\\
        propagation time & T & 1250 &(30.2 fs)\\
        maximal grid extent & $X_{max}$ & $\pm$60 &\\
        absorbing boundary & $x_0$ & $\pm$36 &\\
        \hline
    \end{tabular}
    \caption{Parameters of the calculations in atomic units unless stated otherwise. } 
    \label{table:parameters}
\end{table}

The initial wavefunctions were obtained using imaginary time propagation
filtering out the ground state wavefunction to obtain p$_x$ and p$_y$ orbitals. 
The p$_\pm$ states are defined as p$_\pm=$p$_x\pm i$ p$_y$.
The laser intensity was kept such as not to exceed  5\% ionization and to avoid strong
shifts and mixing of the degenerate atomic orbitals described in \cite{Barth2014}. The spectra were obtained by performing the Fourier transform of the time-dependent dipole acceleration, 
evaluated at every 0.5 a.u.

The results are robust with the variation of the pulse length, the shape and length of 
its rising and falling edges, laser intensity, and wavelength: 
we performed calculations from $\lambda =600$ nm up to $\lambda =1200$ nm.

Fig 1(a) shows reference spectra obtained for 1s initial state of the model potential
with I$_p$ of Neon. It agrees well with previously published results \cite{Long1995,Becker2000a,Kfir2014,Fleischer2014a}, 
the harmonics come in pairs $(n+1)\omega+n 2\omega=
(3n+1)\w$ and $n\omega +(n+1) 2\omega =(3n+2)\w $ of  similar heights. 
The left harmonic in the pair has the same polarization as the fundamental field, 
the right harmonic follows the $2\w$ driver. The harmonics $3n\w$ are parity forbidden.

Figures \ref{fig:spectra}b and \ref{fig:spectra}c show  
spectra for the p$_{+}$ and p$_-$ initial states.
For the p$_+$ initial state, the harmonics that have the same polarization as the driving IR field are preferred. For the p$_-$ initial state, the harmonics with the same polarization as the 2$\w$
driver are stronger. There are additional spectral variations in the plateau region,
different for p$_+$ and p$_-$ orbitals. There is also a qualitative difference between the 
below-threshold (\textless I$_p$) and above-threshold (\textgreater I$_p$) harmonics,
showing that the evolution of the photoelectron in the continuum 
is critical for the observed propensity in the harmonic strengths.

Figure \ref{fig:spectra}d shows the spectra obtained from
adding the contributions from the p$_+$ and p$_-$ orbitals 
coherently, as required. 
In the plateau region, harmonics with the same polarization as 
the driving IR field dominate over those  with opposite polarization.

\begin{figure}[ht!]
\centering
\includegraphics{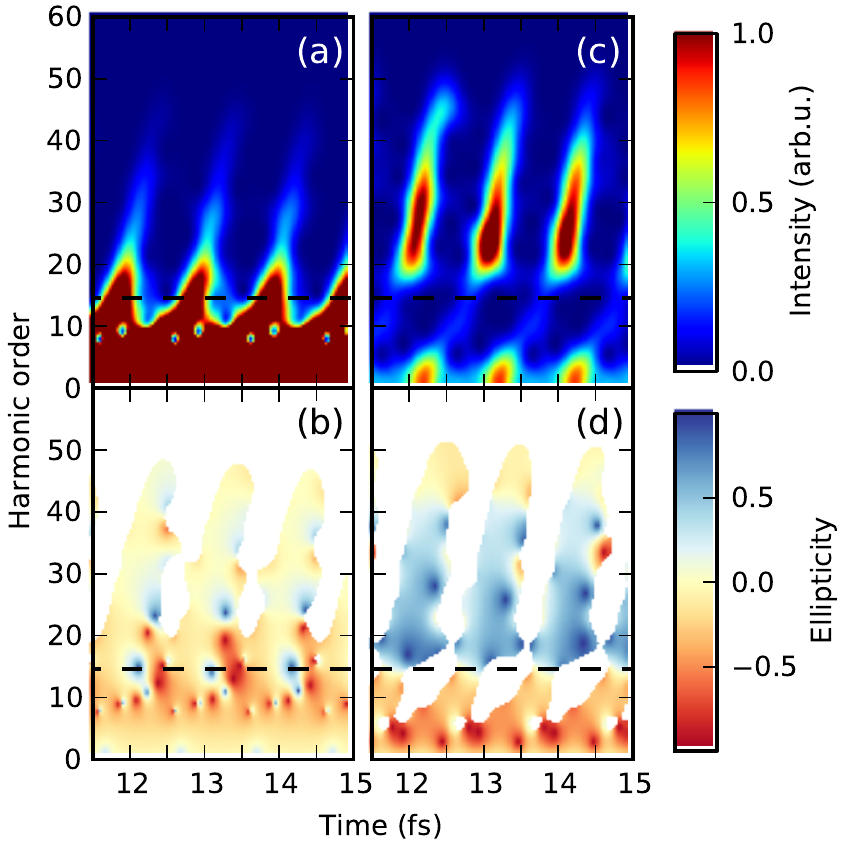}
\caption{Gabor Transformation spectrogram of the harmonic spectra intensity and ellipticity from 1s (a) and (b) and 2p (c) and (d) orbitals. Colour in (b) and (d) indicates the ellipticity of the spectral components in the regions of the spectrogram where the amplitude of the spectra is significant.The horizontal dashed lines mark the I$_p$.}
\label{fig:ellipticity}
\end{figure}

The sub-cycle dynamics of the emission process was analyzed using the Gabor Transform (GT) \cite{Chirila2010} of the time-dependent acceleration dipoles $a(t)$:
\begin{equation}
GT[\Omega, t_0] = \frac{1}{2\pi}\int \mathrm{d}t a(t)\mathrm{e}^{-i \Omega t} \mathrm{e}^{-(t-t_0)^2/(2T^2)}
\end{equation}
where we have chosen  $T=1/3\w$. 
The reference spectrograms for the 1s initial state in Figures \ref{fig:ellipticity}a and \ref{fig:ellipticity}b show the time-dependent intensity (a) and ellipticity (b) for time-resolved
spectra, in the regions where spectral amplitudes are significant. As expected, there are 3 
radiation bursts per $\w$ cycle with linear polarization, as 
predicted in \cite{Long1995,Becker2000a,Becker2000}.

Figures \ref{fig:ellipticity}c and \ref{fig:ellipticity}d show the same spectrogram for the 2p state, 
i.e.  the coherent superposition of the radiation from p$_+$ and p$_-$ states. 
Although the signal strength in the spectrogram is similar to the s orbital, the ellipticity of 
the emitted radiation is very different. Three distinct regions can be identified: (i) below threshold region, where the ellipticity is mostly negative; (ii) the middle region, where the ellipticity is high and positive and (iii) near cutoff region where the emitted radiation is mostly linear. The energy region (ii) of the spectrogram coincides with the spectral window in figure \ref{fig:spectra}d where the  
difference between clockwise and counter-clockwise harmonics is the greatest. 

\begin{figure}[ht!]
\centering
\includegraphics{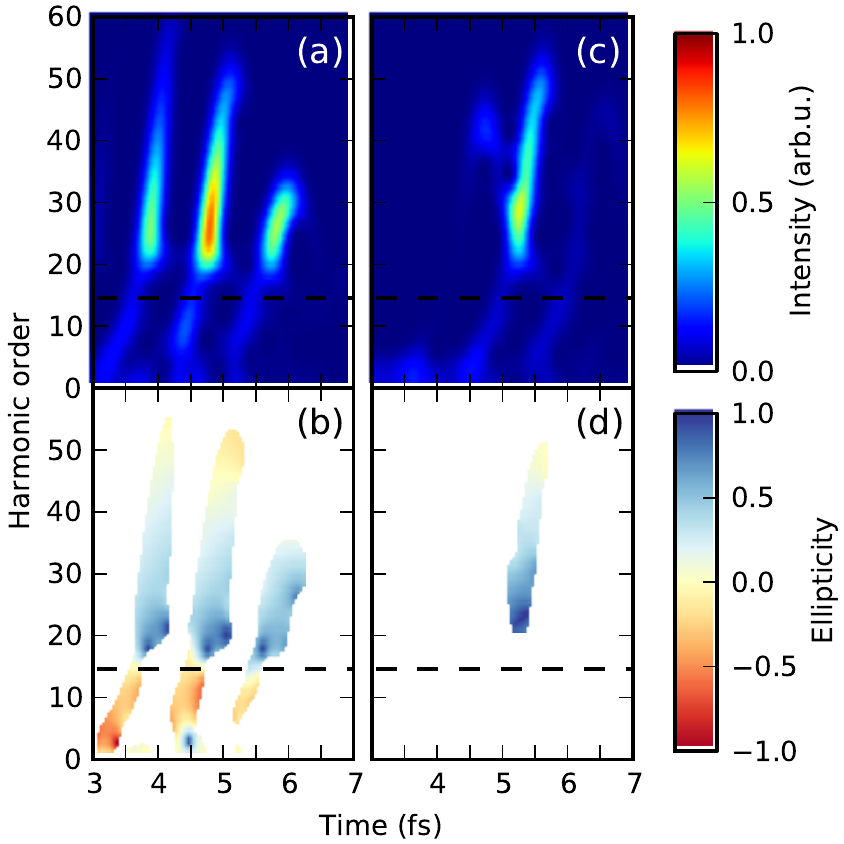}
\caption{Time-resolved XUV emission from a 2p orbital, for time-delayed
8 fs base-to-base 800 nm and 400 nm pulses. (a) Spectral intensity and (b) time-dependent ellipticity
for the perfect overlap of the two-pulses. (c) Spectral intensity and (d) time-dependent ellipticity
for the two-pulse delay of 2 fs shows that single, highly elliptic attosecond pulse 
is generated. 
Colour indicates the ellipticity of the spectral components in the regions of the spectrogram where the amplitude of the spectra is significant.The horizontal dashed lines mark the I$_p$.}
\label{fig:attopulse}
\end{figure}

Application of bi-circular fields 
is naturally extended from the generation 
of an attosecond pulse train to the generation of an isolated
attosecond pulse, using short driving pulses and changing the time-delay
between them. Indeed, the harmonic emission driven by circular fields 
is only possible when the two counter-rotating circular pulses overlap. 
Given high nonlinearity of the overall process (including ionization), 
it will be limited to the temporal window where the  two driving pulses
overlap with nearly equal and high intensity. This idea is tested 
in Fig. \ref{fig:attopulse}, which shows time-resolved spectrograms and
ellipticity of the emitted light for $\lambda=800$ nm and $400$ nm counter-rotating drivers 
with base-to-base duration of 8 fs and $\sin^2$ envelope (3 $\w$ field cycles), for two different
time delays. In  case of perfect overlap, 3 attosecond pulses are generated.
Delaying the low-frequency driving pulse by 2 fs ($\frac{3}{4}$ of $\w$ field period) yields a single
attosecond pulse with highly elliptic polarization.

What is the physical origin of the HHG sensitivity to the angular momentum of the initial state?
The energy and angular momenta that the electron accumulates from the laser field while propagating in the continuum is transferred to the harmonic photon upon recombination. The matrix elements associated with recombination are the complex conjugate of the photoionization matrix elements. In 2D one photon ionization with the field co-rotating with the initial state is much more likely than with counter-rotating field. This is a direct analogue of Fano-Bethe propensity rules \cite{Fano1985}
and is also the case for Rydberg states co-rotating and counter-rotating with the field \cite{Rzazewski1993,Zakrzewski1993}.

Consider the harmonic spectra from p$_+$ orbital.
The right circularly polarized harmonics result from the $(n+1)\w+n2\w$ pathway. The recombination
step is conjugated to photo-ionization from $p_+$ state with a co-rotating field, favoured
by the propensity rules. 
The left-circularly polarized harmonics result from
the  $n\w+(n+1)2\w$ pathway. The recombination
step is conjugated to  photo-ionization from $p_+$ state with a counter-rotating field, dis-favoured
by the propensity rules.  This explains the relative heights of the harmonic pairs for the 
$p_+$ initial state.  The same analysis explains why harmonics co-rotating with $2\w$ field
are preferred for the $p_-$ initial state.

But why is $p_+$ dominant over $p_-$? 
The answer lies in the
stronger effect of the lower-frequency (counter-clockwise) field on the continuum electron, 
which leads to higher population of the continuum states with positive angular momentum
than with the negative one. The more probable recombination from such states is to the $p_+$ state, 
by emitting light with counter-clockwise polarization.

Finally, we comment on the importance of the carrier-envelope phase (CEP) stabilization in this
scheme. As long as the relative phase between the two pulses, $\w$ and $2\w$, is locked,
changing the CEP will rotate the polarization ellipse of the attosecond pulse
but will not alter its high ellipticity. This property, in combination with
the possibility of using relatively routine durations of the two driving pulses, makes
the scheme extremely attractive for practical implementation.

\begin{acknowledgments}
We thank Emilio Pisanty and Felipe Morales for valuable discussions. Financial support 
from the FP7 Marie Curie ITN CORINF, the EPSRC Programme Grant EP/I032517/1, and  
partially from the
US Air Force Office of Scientific Research under
program No. FA9550-12-1-0482 is acknowledged.
\end{acknowledgments}

\bibliography{BicircularHHG}

\begin{thebibliography}{47}%
\makeatletter
\providecommand \@ifxundefined [1]{%
 \@ifx{#1\undefined}
}%
\providecommand \@ifnum [1]{%
 \ifnum #1\expandafter \@firstoftwo
 \else \expandafter \@secondoftwo
 \fi
}%
\providecommand \@ifx [1]{%
 \ifx #1\expandafter \@firstoftwo
 \else \expandafter \@secondoftwo
 \fi
}%
\providecommand \natexlab [1]{#1}%
\providecommand \enquote  [1]{``#1''}%
\providecommand \bibnamefont  [1]{#1}%
\providecommand \bibfnamefont [1]{#1}%
\providecommand \citenamefont [1]{#1}%
\providecommand \href@noop [0]{\@secondoftwo}%
\providecommand \href [0]{\begingroup \@sanitize@url \@href}%
\providecommand \@href[1]{\@@startlink{#1}\@@href}%
\providecommand \@@href[1]{\endgroup#1\@@endlink}%
\providecommand \@sanitize@url [0]{\catcode `\\12\catcode `\$12\catcode
  `\&12\catcode `\#12\catcode `\^12\catcode `\_12\catcode `\%12\relax}%
\providecommand \@@startlink[1]{}%
\providecommand \@@endlink[0]{}%
\providecommand \url  [0]{\begingroup\@sanitize@url \@url }%
\providecommand \@url [1]{\endgroup\@href {#1}{\urlprefix }}%
\providecommand \urlprefix  [0]{URL }%
\providecommand \Eprint [0]{\href }%
\providecommand \doibase [0]{http://dx.doi.org/}%
\providecommand \selectlanguage [0]{\@gobble}%
\providecommand \bibinfo  [0]{\@secondoftwo}%
\providecommand \bibfield  [0]{\@secondoftwo}%
\providecommand \translation [1]{[#1]}%
\providecommand \BibitemOpen [0]{}%
\providecommand \bibitemStop [0]{}%
\providecommand \bibitemNoStop [0]{.\EOS\space}%
\providecommand \EOS [0]{\spacefactor3000\relax}%
\providecommand \BibitemShut  [1]{\csname bibitem#1\endcsname}%
\let\auto@bib@innerbib\@empty
\bibitem [{\citenamefont {Sali\`{e}res}\ \emph
  {et~al.}(1999{\natexlab{a}})\citenamefont {Sali\`{e}res}, \citenamefont
  {L'Huillier}, \citenamefont {Antoine},\ and\ \citenamefont
  {Lewenstein}}]{Salières199983}%
  \BibitemOpen
  \bibfield  {author} {\bibinfo {author} {\bibfnamefont {P.}~\bibnamefont
  {Sali\`{e}res}}, \bibinfo {author} {\bibfnamefont {A.}~\bibnamefont
  {L'Huillier}}, \bibinfo {author} {\bibfnamefont {P.}~\bibnamefont {Antoine}},
  \ and\ \bibinfo {author} {\bibfnamefont {M.}~\bibnamefont {Lewenstein}}\
  }(\bibinfo  {publisher} {Academic Press},\ \bibinfo {year} {1999})\ pp.\
  \bibinfo {pages} {83--142}\BibitemShut {NoStop}%
\bibitem [{\citenamefont {Krausz}\ and\ \citenamefont
  {Ivanov}(2009)}]{Krausz2009}%
  \BibitemOpen
  \bibfield  {author} {\bibinfo {author} {\bibfnamefont {F.}~\bibnamefont
  {Krausz}}\ and\ \bibinfo {author} {\bibfnamefont {M.}~\bibnamefont
  {Ivanov}},\ }\href {\doibase 10.1103/RevModPhys.81.163} {\bibfield  {journal}
  {\bibinfo  {journal} {Reviews of Modern Physics}\ }\textbf {\bibinfo {volume}
  {81}},\ \bibinfo {pages} {163} (\bibinfo {year} {2009})}\BibitemShut
  {NoStop}%
\bibitem [{\citenamefont {Corkum}(1993)}]{Corkum1993}%
  \BibitemOpen
  \bibfield  {author} {\bibinfo {author} {\bibfnamefont {P.~B.}\ \bibnamefont
  {Corkum}},\ }\href@noop {} {\bibfield  {journal} {\bibinfo  {journal}
  {Physical Review}\ }\textbf {\bibinfo {volume} {71}},\ \bibinfo {pages}
  {1994} (\bibinfo {year} {1993})}\BibitemShut {NoStop}%
\bibitem [{\citenamefont {Lewenstein}\ \emph {et~al.}(1994)\citenamefont
  {Lewenstein}, \citenamefont {Balcou}, \citenamefont {Ivanov},\ and\
  \citenamefont {Corkum}}]{Lewenstein1994}%
  \BibitemOpen
  \bibfield  {author} {\bibinfo {author} {\bibfnamefont {M.}~\bibnamefont
  {Lewenstein}}, \bibinfo {author} {\bibfnamefont {P.}~\bibnamefont {Balcou}},
  \bibinfo {author} {\bibfnamefont {M.}~\bibnamefont {Ivanov}}, \ and\ \bibinfo
  {author} {\bibfnamefont {P.}~\bibnamefont {Corkum}},\ }\href
  {http://pra.aps.org/abstract/PRA/v49/i3/p2117_1} {\bibfield  {journal}
  {\bibinfo  {journal} {Physical Review A}\ }\textbf {\bibinfo {volume} {49}},\
  \bibinfo {pages} {2117} (\bibinfo {year} {1994})}\BibitemShut {NoStop}%
\bibitem [{\citenamefont {Haessler}\ \emph {et~al.}(2010)\citenamefont
  {Haessler}, \citenamefont {Caillat}, \citenamefont {Boutu}, \citenamefont
  {Giovanetti-Teixeira}, \citenamefont {Ruchon}, \citenamefont {Auguste},
  \citenamefont {Diveki}, \citenamefont {Breger}, \citenamefont {Maquet},
  \citenamefont {Carr\'{e}}, \citenamefont {Ta\"{\i}eb},\ and\ \citenamefont
  {Sali\`{e}res}}]{Haessler2010}%
  \BibitemOpen
  \bibfield  {author} {\bibinfo {author} {\bibfnamefont {S.}~\bibnamefont
  {Haessler}}, \bibinfo {author} {\bibfnamefont {J.}~\bibnamefont {Caillat}},
  \bibinfo {author} {\bibfnamefont {W.}~\bibnamefont {Boutu}}, \bibinfo
  {author} {\bibfnamefont {C.}~\bibnamefont {Giovanetti-Teixeira}}, \bibinfo
  {author} {\bibfnamefont {T.}~\bibnamefont {Ruchon}}, \bibinfo {author}
  {\bibfnamefont {T.}~\bibnamefont {Auguste}}, \bibinfo {author} {\bibfnamefont
  {Z.}~\bibnamefont {Diveki}}, \bibinfo {author} {\bibfnamefont
  {P.}~\bibnamefont {Breger}}, \bibinfo {author} {\bibfnamefont
  {a.}~\bibnamefont {Maquet}}, \bibinfo {author} {\bibfnamefont
  {B.}~\bibnamefont {Carr\'{e}}}, \bibinfo {author} {\bibfnamefont
  {R.}~\bibnamefont {Ta\"{\i}eb}}, \ and\ \bibinfo {author} {\bibfnamefont
  {P.}~\bibnamefont {Sali\`{e}res}},\ }\href {\doibase 10.1038/nphys1511}
  {\bibfield  {journal} {\bibinfo  {journal} {Nature Physics}\ }\textbf
  {\bibinfo {volume} {6}},\ \bibinfo {pages} {200} (\bibinfo {year}
  {2010})}\BibitemShut {NoStop}%
\bibitem [{\citenamefont {Baker}\ \emph {et~al.}(2006)\citenamefont {Baker},
  \citenamefont {Robinson}, \citenamefont {Haworth}, \citenamefont {Teng},
  \citenamefont {Smith}, \citenamefont {Chirila}, \citenamefont {Lein},
  \citenamefont {Tisch},\ and\ \citenamefont {Marangos}}]{Baker2006b}%
  \BibitemOpen
  \bibfield  {author} {\bibinfo {author} {\bibfnamefont {S.}~\bibnamefont
  {Baker}}, \bibinfo {author} {\bibfnamefont {J.}~\bibnamefont {Robinson}},
  \bibinfo {author} {\bibfnamefont {C.}~\bibnamefont {Haworth}}, \bibinfo
  {author} {\bibfnamefont {H.}~\bibnamefont {Teng}}, \bibinfo {author}
  {\bibfnamefont {R.~A.}\ \bibnamefont {Smith}}, \bibinfo {author}
  {\bibfnamefont {C.~C.}\ \bibnamefont {Chirila}}, \bibinfo {author}
  {\bibfnamefont {M.}~\bibnamefont {Lein}}, \bibinfo {author} {\bibfnamefont
  {J.~W.~G.}\ \bibnamefont {Tisch}}, \ and\ \bibinfo {author} {\bibfnamefont
  {J.~P.}\ \bibnamefont {Marangos}},\ }\href
  {http://www.sciencemag.org/content/312/5772/424.short} {\bibfield  {journal}
  {\bibinfo  {journal} {Science}\ }\textbf {\bibinfo {volume} {229}},\ \bibinfo
  {pages} {424} (\bibinfo {year} {2006})}\BibitemShut {NoStop}%
\bibitem [{\citenamefont {W\"{o}rner}\ \emph {et~al.}(2010)\citenamefont
  {W\"{o}rner}, \citenamefont {Bertrand}, \citenamefont {Kartashov},
  \citenamefont {Corkum},\ and\ \citenamefont {Villeneuve}}]{Worner2010}%
  \BibitemOpen
  \bibfield  {author} {\bibinfo {author} {\bibfnamefont {H.~J.}\ \bibnamefont
  {W\"{o}rner}}, \bibinfo {author} {\bibfnamefont {J.~B.}\ \bibnamefont
  {Bertrand}}, \bibinfo {author} {\bibfnamefont {D.~V.}\ \bibnamefont
  {Kartashov}}, \bibinfo {author} {\bibfnamefont {P.~B.}\ \bibnamefont
  {Corkum}}, \ and\ \bibinfo {author} {\bibfnamefont {D.~M.}\ \bibnamefont
  {Villeneuve}},\ }\href {\doibase 10.1038/nature09185} {\bibfield  {journal}
  {\bibinfo  {journal} {Nature}\ }\textbf {\bibinfo {volume} {466}},\ \bibinfo
  {pages} {604} (\bibinfo {year} {2010})}\BibitemShut {NoStop}%
\bibitem [{\citenamefont {Smirnova}\ \emph {et~al.}(2009)\citenamefont
  {Smirnova}, \citenamefont {Mairesse}, \citenamefont {Patchkovskii},
  \citenamefont {Dudovich}, \citenamefont {Villeneuve}, \citenamefont
  {Corkum},\ and\ \citenamefont {Ivanov}}]{Smirnova2009}%
  \BibitemOpen
  \bibfield  {author} {\bibinfo {author} {\bibfnamefont {O.}~\bibnamefont
  {Smirnova}}, \bibinfo {author} {\bibfnamefont {Y.}~\bibnamefont {Mairesse}},
  \bibinfo {author} {\bibfnamefont {S.}~\bibnamefont {Patchkovskii}}, \bibinfo
  {author} {\bibfnamefont {N.}~\bibnamefont {Dudovich}}, \bibinfo {author}
  {\bibfnamefont {D.}~\bibnamefont {Villeneuve}}, \bibinfo {author}
  {\bibfnamefont {P.}~\bibnamefont {Corkum}}, \ and\ \bibinfo {author}
  {\bibfnamefont {M.~Y.}\ \bibnamefont {Ivanov}},\ }\href {\doibase
  10.1038/nature08253} {\bibfield  {journal} {\bibinfo  {journal} {Nature}\
  }\textbf {\bibinfo {volume} {460}},\ \bibinfo {pages} {972} (\bibinfo {year}
  {2009})}\BibitemShut {NoStop}%
\bibitem [{\citenamefont {Cavalieri}\ \emph {et~al.}(2007)\citenamefont
  {Cavalieri}, \citenamefont {M\"{u}ller}, \citenamefont {Uphues},
  \citenamefont {Yakovlev}, \citenamefont {Baltuska}, \citenamefont {Horvath},
  \citenamefont {Schmidt}, \citenamefont {Bl\"{u}mel}, \citenamefont
  {Holzwarth}, \citenamefont {Hendel}, \citenamefont {Drescher}, \citenamefont
  {Kleineberg}, \citenamefont {Echenique}, \citenamefont {Kienberger},
  \citenamefont {Krausz},\ and\ \citenamefont {Heinzmann}}]{Cavalieri2007}%
  \BibitemOpen
  \bibfield  {author} {\bibinfo {author} {\bibfnamefont {a.~L.}\ \bibnamefont
  {Cavalieri}}, \bibinfo {author} {\bibfnamefont {N.}~\bibnamefont
  {M\"{u}ller}}, \bibinfo {author} {\bibfnamefont {T.}~\bibnamefont {Uphues}},
  \bibinfo {author} {\bibfnamefont {V.~S.}\ \bibnamefont {Yakovlev}}, \bibinfo
  {author} {\bibfnamefont {a.}~\bibnamefont {Baltuska}}, \bibinfo {author}
  {\bibfnamefont {B.}~\bibnamefont {Horvath}}, \bibinfo {author} {\bibfnamefont
  {B.}~\bibnamefont {Schmidt}}, \bibinfo {author} {\bibfnamefont
  {L.}~\bibnamefont {Bl\"{u}mel}}, \bibinfo {author} {\bibfnamefont
  {R.}~\bibnamefont {Holzwarth}}, \bibinfo {author} {\bibfnamefont
  {S.}~\bibnamefont {Hendel}}, \bibinfo {author} {\bibfnamefont
  {M.}~\bibnamefont {Drescher}}, \bibinfo {author} {\bibfnamefont
  {U.}~\bibnamefont {Kleineberg}}, \bibinfo {author} {\bibfnamefont {P.~M.}\
  \bibnamefont {Echenique}}, \bibinfo {author} {\bibfnamefont {R.}~\bibnamefont
  {Kienberger}}, \bibinfo {author} {\bibfnamefont {F.}~\bibnamefont {Krausz}},
  \ and\ \bibinfo {author} {\bibfnamefont {U.}~\bibnamefont {Heinzmann}},\
  }\href {\doibase 10.1038/nature06229} {\bibfield  {journal} {\bibinfo
  {journal} {Nature}\ }\textbf {\bibinfo {volume} {449}},\ \bibinfo {pages}
  {1029} (\bibinfo {year} {2007})}\BibitemShut {NoStop}%
\bibitem [{\citenamefont {Popmintchev}\ \emph {et~al.}(2012)\citenamefont
  {Popmintchev}, \citenamefont {Chen}, \citenamefont {Popmintchev},
  \citenamefont {Paul}, \citenamefont {Brown}, \citenamefont {Ali\v{s}auskas},
  \citenamefont {Andriukaitis}, \citenamefont {Bal\v{c}iūnas}, \citenamefont
  {M\"{u}cke}, \citenamefont {Pugzlys}, \citenamefont {Baltu\v{s}ka},
  \citenamefont {Shim}, \citenamefont {Schrauth}, \citenamefont {Gaeta},
  \citenamefont {Hern\'{a}ndez-Garc\'{\i}a}, \citenamefont {Plaja},
  \citenamefont {Becker}, \citenamefont {Jaron-Becker}, \citenamefont
  {Murnane},\ and\ \citenamefont {Kapteyn}}]{Popmintchev2012}%
  \BibitemOpen
  \bibfield  {author} {\bibinfo {author} {\bibfnamefont {T.}~\bibnamefont
  {Popmintchev}}, \bibinfo {author} {\bibfnamefont {M.~C.}\ \bibnamefont
  {Chen}}, \bibinfo {author} {\bibfnamefont {D.}~\bibnamefont {Popmintchev}},
  \bibinfo {author} {\bibfnamefont {A.}~\bibnamefont {Paul}}, \bibinfo {author}
  {\bibfnamefont {S.}~\bibnamefont {Brown}}, \bibinfo {author} {\bibfnamefont
  {S.}~\bibnamefont {Ali\v{s}auskas}}, \bibinfo {author} {\bibfnamefont
  {G.}~\bibnamefont {Andriukaitis}}, \bibinfo {author} {\bibfnamefont
  {T.}~\bibnamefont {Bal\v{c}iūnas}}, \bibinfo {author} {\bibfnamefont
  {O.~D.}\ \bibnamefont {M\"{u}cke}}, \bibinfo {author} {\bibfnamefont
  {A.}~\bibnamefont {Pugzlys}}, \bibinfo {author} {\bibfnamefont
  {A.}~\bibnamefont {Baltu\v{s}ka}}, \bibinfo {author} {\bibfnamefont
  {B.}~\bibnamefont {Shim}}, \bibinfo {author} {\bibfnamefont {S.}~\bibnamefont
  {Schrauth}}, \bibinfo {author} {\bibfnamefont {A.}~\bibnamefont {Gaeta}},
  \bibinfo {author} {\bibfnamefont {C.}~\bibnamefont
  {Hern\'{a}ndez-Garc\'{\i}a}}, \bibinfo {author} {\bibfnamefont
  {L.}~\bibnamefont {Plaja}}, \bibinfo {author} {\bibfnamefont
  {A.}~\bibnamefont {Becker}}, \bibinfo {author} {\bibfnamefont
  {A.}~\bibnamefont {Jaron-Becker}}, \bibinfo {author} {\bibfnamefont {M.~M.}\
  \bibnamefont {Murnane}}, \ and\ \bibinfo {author} {\bibfnamefont {H.~C.}\
  \bibnamefont {Kapteyn}},\ }\href
  {http://www.sciencemag.org/content/336/6086/1287.short} {\bibfield  {journal}
  {\bibinfo  {journal} {science}\ }\textbf {\bibinfo {volume} {336}},\ \bibinfo
  {pages} {1287} (\bibinfo {year} {2012})}\BibitemShut {NoStop}%
\bibitem [{\citenamefont {Sali\`{e}res}\ \emph
  {et~al.}(1999{\natexlab{b}})\citenamefont {Sali\`{e}res}, \citenamefont
  {D\'{e}roff},\ and\ \citenamefont {Auguste}}]{Salieres1999}%
  \BibitemOpen
  \bibfield  {author} {\bibinfo {author} {\bibfnamefont {P.}~\bibnamefont
  {Sali\`{e}res}}, \bibinfo {author} {\bibfnamefont {L.~L.}\ \bibnamefont
  {D\'{e}roff}}, \ and\ \bibinfo {author} {\bibfnamefont {T.}~\bibnamefont
  {Auguste}},\ }\href
  {http://journals.aps.org/prl/abstract/10.1103/PhysRevLett.83.5483} {\bibfield
   {journal} {\bibinfo  {journal} {Physical Review Letters}\ }\textbf {\bibinfo
  {volume} {83}},\ \bibinfo {pages} {5483} (\bibinfo {year}
  {1999}{\natexlab{b}})}\BibitemShut {NoStop}%
\bibitem [{\citenamefont {L\'{e}pine}\ \emph {et~al.}(2013)\citenamefont
  {L\'{e}pine}, \citenamefont {Sansone},\ and\ \citenamefont
  {Vrakking}}]{Lepine2013}%
  \BibitemOpen
  \bibfield  {author} {\bibinfo {author} {\bibfnamefont {F.}~\bibnamefont
  {L\'{e}pine}}, \bibinfo {author} {\bibfnamefont {G.}~\bibnamefont {Sansone}},
  \ and\ \bibinfo {author} {\bibfnamefont {M.~J.}\ \bibnamefont {Vrakking}},\
  }\href {\doibase 10.1016/j.cplett.2013.05.045} {\bibfield  {journal}
  {\bibinfo  {journal} {Chemical Physics Letters}\ }\textbf {\bibinfo {volume}
  {578}},\ \bibinfo {pages} {1} (\bibinfo {year} {2013})}\BibitemShut {NoStop}%
\bibitem [{\citenamefont {Hergenhahn}\ \emph {et~al.}(2004)\citenamefont
  {Hergenhahn}, \citenamefont {Rennie}, \citenamefont {Kugeler}, \citenamefont
  {Marburger}, \citenamefont {Lischke}, \citenamefont {Powis},\ and\
  \citenamefont {Garcia}}]{Hergenhahn2004}%
  \BibitemOpen
  \bibfield  {author} {\bibinfo {author} {\bibfnamefont {U.}~\bibnamefont
  {Hergenhahn}}, \bibinfo {author} {\bibfnamefont {E.~E.}\ \bibnamefont
  {Rennie}}, \bibinfo {author} {\bibfnamefont {O.}~\bibnamefont {Kugeler}},
  \bibinfo {author} {\bibfnamefont {S.}~\bibnamefont {Marburger}}, \bibinfo
  {author} {\bibfnamefont {T.}~\bibnamefont {Lischke}}, \bibinfo {author}
  {\bibfnamefont {I.}~\bibnamefont {Powis}}, \ and\ \bibinfo {author}
  {\bibfnamefont {G.}~\bibnamefont {Garcia}},\ }\href {\doibase
  10.1063/1.1651474} {\bibfield  {journal} {\bibinfo  {journal} {The Journal of
  chemical physics}\ }\textbf {\bibinfo {volume} {120}},\ \bibinfo {pages}
  {4553} (\bibinfo {year} {2004})}\BibitemShut {NoStop}%
\bibitem [{\citenamefont {B\"{o}wering}\ \emph {et~al.}(2001)\citenamefont
  {B\"{o}wering}, \citenamefont {Lischke}, \citenamefont {Schmidtke},
  \citenamefont {M\"{u}ller}, \citenamefont {Khalil},\ and\ \citenamefont
  {Heinzmann}}]{Bowering2001}%
  \BibitemOpen
  \bibfield  {author} {\bibinfo {author} {\bibfnamefont {N.}~\bibnamefont
  {B\"{o}wering}}, \bibinfo {author} {\bibfnamefont {T.}~\bibnamefont
  {Lischke}}, \bibinfo {author} {\bibfnamefont {B.}~\bibnamefont {Schmidtke}},
  \bibinfo {author} {\bibfnamefont {N.}~\bibnamefont {M\"{u}ller}}, \bibinfo
  {author} {\bibfnamefont {T.}~\bibnamefont {Khalil}}, \ and\ \bibinfo {author}
  {\bibfnamefont {U.}~\bibnamefont {Heinzmann}},\ }\href {\doibase
  10.1103/PhysRevLett.86.1187} {\bibfield  {journal} {\bibinfo  {journal}
  {Physical Review Letters}\ }\textbf {\bibinfo {volume} {86}},\ \bibinfo
  {pages} {1187} (\bibinfo {year} {2001})}\BibitemShut {NoStop}%
\bibitem [{\citenamefont {Powis}(2000)}]{Powis2000}%
  \BibitemOpen
  \bibfield  {author} {\bibinfo {author} {\bibfnamefont {I.}~\bibnamefont
  {Powis}},\ }\href {\doibase 10.1063/1.480581} {\bibfield  {journal} {\bibinfo
   {journal} {The Journal of Chemical Physics}\ }\textbf {\bibinfo {volume}
  {112}},\ \bibinfo {pages} {301} (\bibinfo {year} {2000})}\BibitemShut
  {NoStop}%
\bibitem [{\citenamefont {Ferr\'{e}}\ \emph {et~al.}(2014)\citenamefont
  {Ferr\'{e}}, \citenamefont {Handschin}, \citenamefont {Dumergue},
  \citenamefont {Burgy}, \citenamefont {Comby}, \citenamefont {Descamps},
  \citenamefont {Fabre}, \citenamefont {Garcia}, \citenamefont {G\'{e}neaux},
  \citenamefont {Merceron}, \citenamefont {M\'{e}vel}, \citenamefont {Nahon},
  \citenamefont {Petit}, \citenamefont {Pons}, \citenamefont {Staedter},
  \citenamefont {Weber}, \citenamefont {Ruchon}, \citenamefont {Blanchet},\
  and\ \citenamefont {Mairesse}}]{Ferre2014}%
  \BibitemOpen
  \bibfield  {author} {\bibinfo {author} {\bibfnamefont {A.}~\bibnamefont
  {Ferr\'{e}}}, \bibinfo {author} {\bibfnamefont {C.}~\bibnamefont
  {Handschin}}, \bibinfo {author} {\bibfnamefont {M.}~\bibnamefont {Dumergue}},
  \bibinfo {author} {\bibfnamefont {F.}~\bibnamefont {Burgy}}, \bibinfo
  {author} {\bibfnamefont {A.}~\bibnamefont {Comby}}, \bibinfo {author}
  {\bibfnamefont {D.}~\bibnamefont {Descamps}}, \bibinfo {author}
  {\bibfnamefont {B.}~\bibnamefont {Fabre}}, \bibinfo {author} {\bibfnamefont
  {G.~a.}\ \bibnamefont {Garcia}}, \bibinfo {author} {\bibfnamefont
  {R.}~\bibnamefont {G\'{e}neaux}}, \bibinfo {author} {\bibfnamefont
  {L.}~\bibnamefont {Merceron}}, \bibinfo {author} {\bibfnamefont
  {E.}~\bibnamefont {M\'{e}vel}}, \bibinfo {author} {\bibfnamefont
  {L.}~\bibnamefont {Nahon}}, \bibinfo {author} {\bibfnamefont
  {S.}~\bibnamefont {Petit}}, \bibinfo {author} {\bibfnamefont
  {B.}~\bibnamefont {Pons}}, \bibinfo {author} {\bibfnamefont {D.}~\bibnamefont
  {Staedter}}, \bibinfo {author} {\bibfnamefont {S.}~\bibnamefont {Weber}},
  \bibinfo {author} {\bibfnamefont {T.}~\bibnamefont {Ruchon}}, \bibinfo
  {author} {\bibfnamefont {V.}~\bibnamefont {Blanchet}}, \ and\ \bibinfo
  {author} {\bibfnamefont {Y.}~\bibnamefont {Mairesse}},\ }\href {\doibase
  10.1038/nphoton.2014.314} {\bibfield  {journal} {\bibinfo  {journal} {Nature
  Photonics}\ }\textbf {\bibinfo {volume} {9}},\ \bibinfo {pages} {93}
  (\bibinfo {year} {2014})}\BibitemShut {NoStop}%
\bibitem [{\citenamefont {Travnikova}\ \emph {et~al.}(2010)\citenamefont
  {Travnikova}, \citenamefont {Liu}, \citenamefont {Lindblad}, \citenamefont
  {Nicolas}, \citenamefont {S\"{o}derstr\"{o}m}, \citenamefont {Kimberg},
  \citenamefont {Gel’mukhanov},\ and\ \citenamefont
  {Miron}}]{Travnikova2010}%
  \BibitemOpen
  \bibfield  {author} {\bibinfo {author} {\bibfnamefont {O.}~\bibnamefont
  {Travnikova}}, \bibinfo {author} {\bibfnamefont {J.-C.}\ \bibnamefont {Liu}},
  \bibinfo {author} {\bibfnamefont {A.}~\bibnamefont {Lindblad}}, \bibinfo
  {author} {\bibfnamefont {C.}~\bibnamefont {Nicolas}}, \bibinfo {author}
  {\bibfnamefont {J.}~\bibnamefont {S\"{o}derstr\"{o}m}}, \bibinfo {author}
  {\bibfnamefont {V.}~\bibnamefont {Kimberg}}, \bibinfo {author} {\bibfnamefont
  {F.}~\bibnamefont {Gel’mukhanov}}, \ and\ \bibinfo {author} {\bibfnamefont
  {C.}~\bibnamefont {Miron}},\ }\href {\doibase 10.1103/PhysRevLett.105.233001}
  {\bibfield  {journal} {\bibinfo  {journal} {Physical Review Letters}\
  }\textbf {\bibinfo {volume} {105}},\ \bibinfo {pages} {233001} (\bibinfo
  {year} {2010})}\BibitemShut {NoStop}%
\bibitem [{\citenamefont {Radu}\ \emph {et~al.}(2011)\citenamefont {Radu},
  \citenamefont {Vahaplar}, \citenamefont {Stamm}, \citenamefont {Kachel},
  \citenamefont {Pontius}, \citenamefont {D\"{u}rr}, \citenamefont {Ostler},
  \citenamefont {Barker}, \citenamefont {Evans}, \citenamefont {Chantrell},
  \citenamefont {Tsukamoto}, \citenamefont {Itoh}, \citenamefont {Kirilyuk},
  \citenamefont {Rasing},\ and\ \citenamefont {Kimel}}]{Radu2011}%
  \BibitemOpen
  \bibfield  {author} {\bibinfo {author} {\bibfnamefont {I.}~\bibnamefont
  {Radu}}, \bibinfo {author} {\bibfnamefont {K.}~\bibnamefont {Vahaplar}},
  \bibinfo {author} {\bibfnamefont {C.}~\bibnamefont {Stamm}}, \bibinfo
  {author} {\bibfnamefont {T.}~\bibnamefont {Kachel}}, \bibinfo {author}
  {\bibfnamefont {N.}~\bibnamefont {Pontius}}, \bibinfo {author} {\bibfnamefont
  {H.~a.}\ \bibnamefont {D\"{u}rr}}, \bibinfo {author} {\bibfnamefont {T.~a.}\
  \bibnamefont {Ostler}}, \bibinfo {author} {\bibfnamefont {J.}~\bibnamefont
  {Barker}}, \bibinfo {author} {\bibfnamefont {R.~F.~L.}\ \bibnamefont
  {Evans}}, \bibinfo {author} {\bibfnamefont {R.~W.}\ \bibnamefont
  {Chantrell}}, \bibinfo {author} {\bibfnamefont {a.}~\bibnamefont
  {Tsukamoto}}, \bibinfo {author} {\bibfnamefont {a.}~\bibnamefont {Itoh}},
  \bibinfo {author} {\bibfnamefont {a.}~\bibnamefont {Kirilyuk}}, \bibinfo
  {author} {\bibfnamefont {T.}~\bibnamefont {Rasing}}, \ and\ \bibinfo {author}
  {\bibfnamefont {a.~V.}\ \bibnamefont {Kimel}},\ }\href {\doibase
  10.1038/nature09901} {\bibfield  {journal} {\bibinfo  {journal} {Nature}\
  }\textbf {\bibinfo {volume} {472}},\ \bibinfo {pages} {205} (\bibinfo {year}
  {2011})}\BibitemShut {NoStop}%
\bibitem [{\citenamefont {Boeglin}\ \emph {et~al.}(2010)\citenamefont
  {Boeglin}, \citenamefont {Beaurepaire}, \citenamefont {Halt\'{e}},
  \citenamefont {L\'{o}pez-Flores}, \citenamefont {Stamm}, \citenamefont
  {Pontius}, \citenamefont {D\"{u}rr},\ and\ \citenamefont
  {Bigot}}]{Boeglin2010}%
  \BibitemOpen
  \bibfield  {author} {\bibinfo {author} {\bibfnamefont {C.}~\bibnamefont
  {Boeglin}}, \bibinfo {author} {\bibfnamefont {E.}~\bibnamefont
  {Beaurepaire}}, \bibinfo {author} {\bibfnamefont {V.}~\bibnamefont
  {Halt\'{e}}}, \bibinfo {author} {\bibfnamefont {V.}~\bibnamefont
  {L\'{o}pez-Flores}}, \bibinfo {author} {\bibfnamefont {C.}~\bibnamefont
  {Stamm}}, \bibinfo {author} {\bibfnamefont {N.}~\bibnamefont {Pontius}},
  \bibinfo {author} {\bibfnamefont {H.~a.}\ \bibnamefont {D\"{u}rr}}, \ and\
  \bibinfo {author} {\bibfnamefont {J.-Y.}\ \bibnamefont {Bigot}},\ }\href
  {\doibase 10.1038/nature09070} {\bibfield  {journal} {\bibinfo  {journal}
  {Nature}\ }\textbf {\bibinfo {volume} {465}},\ \bibinfo {pages} {458}
  (\bibinfo {year} {2010})}\BibitemShut {NoStop}%
\bibitem [{\citenamefont {Eisebitt}\ \emph {et~al.}(2004)\citenamefont
  {Eisebitt}, \citenamefont {L\"{u}ning}, \citenamefont {Schlotter},\ and\
  \citenamefont {L\"{o}rgen}}]{Eisebitt2004}%
  \BibitemOpen
  \bibfield  {author} {\bibinfo {author} {\bibfnamefont {S.}~\bibnamefont
  {Eisebitt}}, \bibinfo {author} {\bibfnamefont {J.}~\bibnamefont
  {L\"{u}ning}}, \bibinfo {author} {\bibfnamefont {W.}~\bibnamefont
  {Schlotter}}, \ and\ \bibinfo {author} {\bibfnamefont {M.}~\bibnamefont
  {L\"{o}rgen}},\ }\href {\doibase 10.1038/nature03129.1.} {\bibfield
  {journal} {\bibinfo  {journal} {Nature}\ }\textbf {\bibinfo {volume} {432}},\
  \bibinfo {pages} {885} (\bibinfo {year} {2004})}\BibitemShut {NoStop}%
\bibitem [{\citenamefont {Fischer}\ \emph {et~al.}(1999)\citenamefont
  {Fischer}, \citenamefont {Eim\"{u}ller}, \citenamefont {Sch\"{u}tz},
  \citenamefont {Schmahl}, \citenamefont {Guttmann},\ and\ \citenamefont
  {Bayreuther}}]{Fischer1999}%
  \BibitemOpen
  \bibfield  {author} {\bibinfo {author} {\bibfnamefont {P.}~\bibnamefont
  {Fischer}}, \bibinfo {author} {\bibfnamefont {T.}~\bibnamefont
  {Eim\"{u}ller}}, \bibinfo {author} {\bibfnamefont {G.}~\bibnamefont
  {Sch\"{u}tz}}, \bibinfo {author} {\bibfnamefont {G.}~\bibnamefont {Schmahl}},
  \bibinfo {author} {\bibfnamefont {P.}~\bibnamefont {Guttmann}}, \ and\
  \bibinfo {author} {\bibfnamefont {G.}~\bibnamefont {Bayreuther}},\ }\href
  {http://www.sciencedirect.com/science/article/pii/S0304885398012104}
  {\bibfield  {journal} {\bibinfo  {journal} {Journal of Magnetism and Magnetic
  Materials}\ }\textbf {\bibinfo {volume} {198-199}},\ \bibinfo {pages} {624}
  (\bibinfo {year} {1999})}\BibitemShut {NoStop}%
\bibitem [{\citenamefont {L\'{o}pez-Flores}\ \emph {et~al.}(2012)\citenamefont
  {L\'{o}pez-Flores}, \citenamefont {Arabski}, \citenamefont {Stamm},
  \citenamefont {Halt\'{e}}, \citenamefont {Pontius}, \citenamefont
  {Beaurepaire},\ and\ \citenamefont {Boeglin}}]{Lopez-Flores2012}%
  \BibitemOpen
  \bibfield  {author} {\bibinfo {author} {\bibfnamefont {V.}~\bibnamefont
  {L\'{o}pez-Flores}}, \bibinfo {author} {\bibfnamefont {J.}~\bibnamefont
  {Arabski}}, \bibinfo {author} {\bibfnamefont {C.}~\bibnamefont {Stamm}},
  \bibinfo {author} {\bibfnamefont {V.}~\bibnamefont {Halt\'{e}}}, \bibinfo
  {author} {\bibfnamefont {N.}~\bibnamefont {Pontius}}, \bibinfo {author}
  {\bibfnamefont {E.}~\bibnamefont {Beaurepaire}}, \ and\ \bibinfo {author}
  {\bibfnamefont {C.}~\bibnamefont {Boeglin}},\ }\href {\doibase
  10.1103/PhysRevB.86.014424} {\bibfield  {journal} {\bibinfo  {journal}
  {Physical Review B}\ }\textbf {\bibinfo {volume} {86}},\ \bibinfo {pages}
  {014424} (\bibinfo {year} {2012})}\BibitemShut {NoStop}%
\bibitem [{\citenamefont {St\"{o}hr}\ \emph {et~al.}(1993)\citenamefont
  {St\"{o}hr}, \citenamefont {Wu}, \citenamefont {Hermsmeier}, \citenamefont
  {Samant}, \citenamefont {Harp}, \citenamefont {Koranda}, \citenamefont
  {Dunham},\ and\ \citenamefont {Tonner}}]{Stohr1993}%
  \BibitemOpen
  \bibfield  {author} {\bibinfo {author} {\bibfnamefont {J.}~\bibnamefont
  {St\"{o}hr}}, \bibinfo {author} {\bibfnamefont {Y.}~\bibnamefont {Wu}},
  \bibinfo {author} {\bibfnamefont {B.~D.}\ \bibnamefont {Hermsmeier}},
  \bibinfo {author} {\bibfnamefont {M.~G.}\ \bibnamefont {Samant}}, \bibinfo
  {author} {\bibfnamefont {G.~R.}\ \bibnamefont {Harp}}, \bibinfo {author}
  {\bibfnamefont {S.}~\bibnamefont {Koranda}}, \bibinfo {author} {\bibfnamefont
  {D.}~\bibnamefont {Dunham}}, \ and\ \bibinfo {author} {\bibfnamefont {B.~P.}\
  \bibnamefont {Tonner}},\ }\href
  {https://www-ssrl.slac.stanford.edu/stohr/science-259-658-93.pdf} {\bibfield
  {journal} {\bibinfo  {journal} {Science}\ }\textbf {\bibinfo {volume}
  {259}},\ \bibinfo {pages} {658} (\bibinfo {year} {1993})}\BibitemShut
  {NoStop}%
\bibitem [{\citenamefont {Sch\"{u}tz}\ \emph {et~al.}(1993)\citenamefont
  {Sch\"{u}tz}, \citenamefont {Kn\"{u}lle},\ and\ \citenamefont
  {Ebert}}]{Schutz1993}%
  \BibitemOpen
  \bibfield  {author} {\bibinfo {author} {\bibfnamefont {G.}~\bibnamefont
  {Sch\"{u}tz}}, \bibinfo {author} {\bibfnamefont {M.}~\bibnamefont
  {Kn\"{u}lle}}, \ and\ \bibinfo {author} {\bibfnamefont {H.}~\bibnamefont
  {Ebert}},\ }\href {http://iopscience.iop.org/1402-4896/1993/T49A/053}
  {\bibfield  {journal} {\bibinfo  {journal} {Physica Scripta}\ }\textbf
  {\bibinfo {volume} {302}} (\bibinfo {year} {1993})}\BibitemShut {NoStop}%
\bibitem [{\citenamefont {Chen}\ \emph {et~al.}(2015)\citenamefont {Chen},
  \citenamefont {Luo},\ and\ \citenamefont {Luo}}]{Chen2015}%
  \BibitemOpen
  \bibfield  {author} {\bibinfo {author} {\bibfnamefont {F.}~\bibnamefont
  {Chen}}, \bibinfo {author} {\bibfnamefont {J.}~\bibnamefont {Luo}}, \ and\
  \bibinfo {author} {\bibfnamefont {F.}~\bibnamefont {Luo}},\ }\href {\doibase
  10.1016/j.optcom.2014.12.034} {\bibfield  {journal} {\bibinfo  {journal}
  {Optics Communications}\ }\textbf {\bibinfo {volume} {342}},\ \bibinfo
  {pages} {68} (\bibinfo {year} {2015})}\BibitemShut {NoStop}%
\bibitem [{\citenamefont {Morales}\ \emph {et~al.}(2012)\citenamefont
  {Morales}, \citenamefont {Barth}, \citenamefont {Serbinenko}, \citenamefont
  {Patchkovskii},\ and\ \citenamefont {Smirnova}}]{Morales2012}%
  \BibitemOpen
  \bibfield  {author} {\bibinfo {author} {\bibfnamefont {F.}~\bibnamefont
  {Morales}}, \bibinfo {author} {\bibfnamefont {I.}~\bibnamefont {Barth}},
  \bibinfo {author} {\bibfnamefont {V.}~\bibnamefont {Serbinenko}}, \bibinfo
  {author} {\bibfnamefont {S.}~\bibnamefont {Patchkovskii}}, \ and\ \bibinfo
  {author} {\bibfnamefont {O.}~\bibnamefont {Smirnova}},\ }\href {\doibase
  10.1080/09500340.2012.690051} {\bibfield  {journal} {\bibinfo  {journal}
  {Journal of Modern Optics}\ }\textbf {\bibinfo {volume} {59}},\ \bibinfo
  {pages} {1303} (\bibinfo {year} {2012})}\BibitemShut {NoStop}%
\bibitem [{\citenamefont {Yuan}\ and\ \citenamefont
  {Bandrauk}(2013)}]{Yuan2013}%
  \BibitemOpen
  \bibfield  {author} {\bibinfo {author} {\bibfnamefont {K.-J.}\ \bibnamefont
  {Yuan}}\ and\ \bibinfo {author} {\bibfnamefont {A.}~\bibnamefont
  {Bandrauk}},\ }\href {\doibase 10.1103/PhysRevLett.110.023003} {\bibfield
  {journal} {\bibinfo  {journal} {Physical Review Letters}\ }\textbf {\bibinfo
  {volume} {110}},\ \bibinfo {pages} {023003} (\bibinfo {year}
  {2013})}\BibitemShut {NoStop}%
\bibitem [{\citenamefont {Milo\v{s}evi\'{c}}\ \emph {et~al.}(2000)\citenamefont
  {Milo\v{s}evi\'{c}}, \citenamefont {Becker},\ and\ \citenamefont
  {Kopold}}]{Becker2000a}%
  \BibitemOpen
  \bibfield  {author} {\bibinfo {author} {\bibfnamefont {D.~B.}\ \bibnamefont
  {Milo\v{s}evi\'{c}}}, \bibinfo {author} {\bibfnamefont {W.}~\bibnamefont
  {Becker}}, \ and\ \bibinfo {author} {\bibfnamefont {R.}~\bibnamefont
  {Kopold}},\ }\href {\doibase 10.1103/PhysRevA.61.063403} {\bibfield
  {journal} {\bibinfo  {journal} {Physical Review A}\ }\textbf {\bibinfo
  {volume} {61}},\ \bibinfo {pages} {063403} (\bibinfo {year}
  {2000})}\BibitemShut {NoStop}%
\bibitem [{\citenamefont {Milo\v{s}evi\'{c}}\ and\ \citenamefont
  {Becker}(2000)}]{Becker2000}%
  \BibitemOpen
  \bibfield  {author} {\bibinfo {author} {\bibfnamefont {D.}~\bibnamefont
  {Milo\v{s}evi\'{c}}}\ and\ \bibinfo {author} {\bibfnamefont {W.}~\bibnamefont
  {Becker}},\ }\href {\doibase 10.1103/PhysRevA.62.011403} {\bibfield
  {journal} {\bibinfo  {journal} {Physical Review A}\ }\textbf {\bibinfo
  {volume} {62}},\ \bibinfo {pages} {011403} (\bibinfo {year}
  {2000})}\BibitemShut {NoStop}%
\bibitem [{\citenamefont {Yuan}\ and\ \citenamefont
  {Bandrauk}(2011)}]{Yuan2011}%
  \BibitemOpen
  \bibfield  {author} {\bibinfo {author} {\bibfnamefont {K.-J.}\ \bibnamefont
  {Yuan}}\ and\ \bibinfo {author} {\bibfnamefont {A.~D.}\ \bibnamefont
  {Bandrauk}},\ }\href {\doibase 10.1103/PhysRevA.84.023410} {\bibfield
  {journal} {\bibinfo  {journal} {Physical Review A}\ }\textbf {\bibinfo
  {volume} {84}},\ \bibinfo {pages} {023410} (\bibinfo {year}
  {2011})}\BibitemShut {NoStop}%
\bibitem [{\citenamefont {Eichmann}\ \emph {et~al.}(1995)\citenamefont
  {Eichmann}, \citenamefont {Egbert}, \citenamefont {Nolte}, \citenamefont
  {Momma},\ and\ \citenamefont {Wellegehausen}}]{Eichmann1995}%
  \BibitemOpen
  \bibfield  {author} {\bibinfo {author} {\bibfnamefont {H.}~\bibnamefont
  {Eichmann}}, \bibinfo {author} {\bibfnamefont {A.}~\bibnamefont {Egbert}},
  \bibinfo {author} {\bibfnamefont {S.}~\bibnamefont {Nolte}}, \bibinfo
  {author} {\bibfnamefont {C.}~\bibnamefont {Momma}}, \ and\ \bibinfo {author}
  {\bibfnamefont {B.}~\bibnamefont {Wellegehausen}},\ }\href {\doibase
  http://dx.doi.org/10.1103/PhysRevA.51.R3414} {\bibfield  {journal} {\bibinfo
  {journal} {Physical Review A}\ }\textbf {\bibinfo {volume} {51}},\ \bibinfo
  {pages} {3414} (\bibinfo {year} {1995})}\BibitemShut {NoStop}%
\bibitem [{\citenamefont {Long}\ \emph {et~al.}(1995)\citenamefont {Long},
  \citenamefont {Becker},\ and\ \citenamefont {McIver}}]{Long1995}%
  \BibitemOpen
  \bibfield  {author} {\bibinfo {author} {\bibfnamefont {S.}~\bibnamefont
  {Long}}, \bibinfo {author} {\bibfnamefont {W.}~\bibnamefont {Becker}}, \ and\
  \bibinfo {author} {\bibfnamefont {J.~K.}\ \bibnamefont {McIver}},\ }\href
  {\doibase http://dx.doi.org/10.1103/PhysRevA.51.R3414} {\bibfield  {journal}
  {\bibinfo  {journal} {Physical Review A}\ }\textbf {\bibinfo {volume} {52}},\
  \bibinfo {pages} {2262} (\bibinfo {year} {1995})}\BibitemShut {NoStop}%
\bibitem [{\citenamefont {Pisanty}\ \emph {et~al.}(2014)\citenamefont
  {Pisanty}, \citenamefont {Sukiasyan},\ and\ \citenamefont
  {Ivanov}}]{Pisanty2014}%
  \BibitemOpen
  \bibfield  {author} {\bibinfo {author} {\bibfnamefont {E.}~\bibnamefont
  {Pisanty}}, \bibinfo {author} {\bibfnamefont {S.}~\bibnamefont {Sukiasyan}},
  \ and\ \bibinfo {author} {\bibfnamefont {M.}~\bibnamefont {Ivanov}},\ }\href
  {\doibase 10.1103/PhysRevA.90.043829} {\bibfield  {journal} {\bibinfo
  {journal} {Physical Review A}\ }\textbf {\bibinfo {volume} {90}},\ \bibinfo
  {pages} {043829} (\bibinfo {year} {2014})}\BibitemShut {NoStop}%
\bibitem [{\citenamefont {Ivanov}\ and\ \citenamefont
  {Pisanty}(2014)}]{Ivanov2014}%
  \BibitemOpen
  \bibfield  {author} {\bibinfo {author} {\bibfnamefont {M.}~\bibnamefont
  {Ivanov}}\ and\ \bibinfo {author} {\bibfnamefont {E.}~\bibnamefont
  {Pisanty}},\ }\href {\doibase 10.1038/nphoton.2014.141} {\bibfield  {journal}
  {\bibinfo  {journal} {Nature Photonics}\ }\textbf {\bibinfo {volume} {8}},\
  \bibinfo {pages} {501} (\bibinfo {year} {2014})}\BibitemShut {NoStop}%
\bibitem [{\citenamefont {Fleischer}\ \emph {et~al.}(2014)\citenamefont
  {Fleischer}, \citenamefont {Kfir}, \citenamefont {Diskin}, \citenamefont
  {Sidorenko},\ and\ \citenamefont {Cohen}}]{Fleischer2014a}%
  \BibitemOpen
  \bibfield  {author} {\bibinfo {author} {\bibfnamefont {A.}~\bibnamefont
  {Fleischer}}, \bibinfo {author} {\bibfnamefont {O.}~\bibnamefont {Kfir}},
  \bibinfo {author} {\bibfnamefont {T.}~\bibnamefont {Diskin}}, \bibinfo
  {author} {\bibfnamefont {P.}~\bibnamefont {Sidorenko}}, \ and\ \bibinfo
  {author} {\bibfnamefont {O.}~\bibnamefont {Cohen}},\ }\href {\doibase
  10.1038/nphoton.2014.108} {\bibfield  {journal} {\bibinfo  {journal} {Nature
  Photonics}\ }\textbf {\bibinfo {volume} {8}},\ \bibinfo {pages} {543}
  (\bibinfo {year} {2014})}\BibitemShut {NoStop}%
\bibitem [{\citenamefont {Kfir}\ \emph {et~al.}(2014)\citenamefont {Kfir},
  \citenamefont {Grychtol}, \citenamefont {Turgut}, \citenamefont {Knut},
  \citenamefont {Zusin}, \citenamefont {Popmintchev}, \citenamefont
  {Popmintchev}, \citenamefont {Nembach}, \citenamefont {Shaw}, \citenamefont
  {Fleischer}, \citenamefont {Kapteyn}, \citenamefont {Murnane},\ and\
  \citenamefont {Cohen}}]{Kfir2014}%
  \BibitemOpen
  \bibfield  {author} {\bibinfo {author} {\bibfnamefont {O.}~\bibnamefont
  {Kfir}}, \bibinfo {author} {\bibfnamefont {P.}~\bibnamefont {Grychtol}},
  \bibinfo {author} {\bibfnamefont {E.}~\bibnamefont {Turgut}}, \bibinfo
  {author} {\bibfnamefont {R.}~\bibnamefont {Knut}}, \bibinfo {author}
  {\bibfnamefont {D.}~\bibnamefont {Zusin}}, \bibinfo {author} {\bibfnamefont
  {D.}~\bibnamefont {Popmintchev}}, \bibinfo {author} {\bibfnamefont
  {T.}~\bibnamefont {Popmintchev}}, \bibinfo {author} {\bibfnamefont
  {H.}~\bibnamefont {Nembach}}, \bibinfo {author} {\bibfnamefont {J.~M.}\
  \bibnamefont {Shaw}}, \bibinfo {author} {\bibfnamefont {A.}~\bibnamefont
  {Fleischer}}, \bibinfo {author} {\bibfnamefont {H.}~\bibnamefont {Kapteyn}},
  \bibinfo {author} {\bibfnamefont {M.}~\bibnamefont {Murnane}}, \ and\
  \bibinfo {author} {\bibfnamefont {O.}~\bibnamefont {Cohen}},\ }\href
  {\doibase 10.1038/nphoton.2014.293} {\bibfield  {journal} {\bibinfo
  {journal} {Nature Photonics}\ }\textbf {\bibinfo {volume} {9}},\ \bibinfo
  {pages} {99} (\bibinfo {year} {2014})}\BibitemShut {NoStop}%
\bibitem [{\citenamefont {Becker}\ \emph {et~al.}(1999)\citenamefont {Becker},
  \citenamefont {Chichkov},\ and\ \citenamefont {Wellegehausen}}]{Becker1999}%
  \BibitemOpen
  \bibfield  {author} {\bibinfo {author} {\bibfnamefont {W.}~\bibnamefont
  {Becker}}, \bibinfo {author} {\bibfnamefont {B.}~\bibnamefont {Chichkov}}, \
  and\ \bibinfo {author} {\bibfnamefont {B.}~\bibnamefont {Wellegehausen}},\
  }\href {http://journals.aps.org/pra/abstract/10.1103/PhysRevA.60.1721}
  {\bibfield  {journal} {\bibinfo  {journal} {Physical Review A}\ }\textbf
  {\bibinfo {volume} {60}},\ \bibinfo {pages} {1721} (\bibinfo {year}
  {1999})}\BibitemShut {NoStop}%
\bibitem [{\citenamefont {Yuan}\ and\ \citenamefont
  {Bandrauk}(2012)}]{Yuan2012a}%
  \BibitemOpen
  \bibfield  {author} {\bibinfo {author} {\bibfnamefont {K.-J.}\ \bibnamefont
  {Yuan}}\ and\ \bibinfo {author} {\bibfnamefont {A.~D.}\ \bibnamefont
  {Bandrauk}},\ }\href {\doibase 10.1088/0953-4075/45/7/074001} {\bibfield
  {journal} {\bibinfo  {journal} {Journal of Physics B: Atomic, Molecular and
  Optical Physics}\ }\textbf {\bibinfo {volume} {45}},\ \bibinfo {pages}
  {074001} (\bibinfo {year} {2012})}\BibitemShut {NoStop}%
\bibitem [{\citenamefont {Lambert}\ \emph {et~al.}(2015)\citenamefont
  {Lambert}, \citenamefont {Vodungbo}, \citenamefont {Gautier}, \citenamefont
  {Mahieu}, \citenamefont {Malka}, \citenamefont {Sebban}, \citenamefont
  {Zeitoun}, \citenamefont {Luning}, \citenamefont {Perron}, \citenamefont
  {Andreev}, \citenamefont {Stremoukhov}, \citenamefont {Ardana-Lamas},
  \citenamefont {Dax}, \citenamefont {Hauri}, \citenamefont {Sardinha},\ and\
  \citenamefont {Fajardo}}]{Lambert2015}%
  \BibitemOpen
  \bibfield  {author} {\bibinfo {author} {\bibfnamefont {G.}~\bibnamefont
  {Lambert}}, \bibinfo {author} {\bibfnamefont {B.}~\bibnamefont {Vodungbo}},
  \bibinfo {author} {\bibfnamefont {J.}~\bibnamefont {Gautier}}, \bibinfo
  {author} {\bibfnamefont {B.}~\bibnamefont {Mahieu}}, \bibinfo {author}
  {\bibfnamefont {V.}~\bibnamefont {Malka}}, \bibinfo {author} {\bibfnamefont
  {S.}~\bibnamefont {Sebban}}, \bibinfo {author} {\bibfnamefont
  {P.}~\bibnamefont {Zeitoun}}, \bibinfo {author} {\bibfnamefont
  {J.}~\bibnamefont {Luning}}, \bibinfo {author} {\bibfnamefont
  {J.}~\bibnamefont {Perron}}, \bibinfo {author} {\bibfnamefont
  {a.}~\bibnamefont {Andreev}}, \bibinfo {author} {\bibfnamefont
  {S.}~\bibnamefont {Stremoukhov}}, \bibinfo {author} {\bibfnamefont
  {F.}~\bibnamefont {Ardana-Lamas}}, \bibinfo {author} {\bibfnamefont
  {a.}~\bibnamefont {Dax}}, \bibinfo {author} {\bibfnamefont {C.~P.}\
  \bibnamefont {Hauri}}, \bibinfo {author} {\bibfnamefont {a.}~\bibnamefont
  {Sardinha}}, \ and\ \bibinfo {author} {\bibfnamefont {M.}~\bibnamefont
  {Fajardo}},\ }\href {\doibase 10.1038/ncomms7167} {\bibfield  {journal}
  {\bibinfo  {journal} {Nature communications}\ }\textbf {\bibinfo {volume}
  {6}},\ \bibinfo {pages} {6167} (\bibinfo {year} {2015})}\BibitemShut
  {NoStop}%
\bibitem [{\citenamefont {Alon}\ \emph {et~al.}(1998)\citenamefont {Alon},
  \citenamefont {Averbukh},\ and\ \citenamefont {Moiseyev}}]{Alon1998}%
  \BibitemOpen
  \bibfield  {author} {\bibinfo {author} {\bibfnamefont {O.}~\bibnamefont
  {Alon}}, \bibinfo {author} {\bibfnamefont {V.}~\bibnamefont {Averbukh}}, \
  and\ \bibinfo {author} {\bibfnamefont {N.}~\bibnamefont {Moiseyev}},\ }\href
  {http://journals.aps.org/prl/abstract/10.1103/PhysRevLett.80.3743} {\bibfield
   {journal} {\bibinfo  {journal} {Physical Review Letters}\ }\textbf {\bibinfo
  {volume} {80}},\ \bibinfo {pages} {3743} (\bibinfo {year}
  {1998})}\BibitemShut {NoStop}%
\bibitem [{\citenamefont {Mauger}\ \emph {et~al.}(2014)\citenamefont {Mauger},
  \citenamefont {Bandrauk}, \citenamefont {Kamor}, \citenamefont {Uzer},\ and\
  \citenamefont {Chandre}}]{Mauger2014}%
  \BibitemOpen
  \bibfield  {author} {\bibinfo {author} {\bibfnamefont {F.}~\bibnamefont
  {Mauger}}, \bibinfo {author} {\bibfnamefont {a.~D.}\ \bibnamefont
  {Bandrauk}}, \bibinfo {author} {\bibfnamefont {a.}~\bibnamefont {Kamor}},
  \bibinfo {author} {\bibfnamefont {T.}~\bibnamefont {Uzer}}, \ and\ \bibinfo
  {author} {\bibfnamefont {C.}~\bibnamefont {Chandre}},\ }\href {\doibase
  10.1088/0953-4075/47/4/041001} {\bibfield  {journal} {\bibinfo  {journal}
  {Journal of Physics B: Atomic, Molecular and Optical Physics}\ }\textbf
  {\bibinfo {volume} {47}},\ \bibinfo {pages} {041001} (\bibinfo {year}
  {2014})}\BibitemShut {NoStop}%
\bibitem [{\citenamefont {Barth}\ and\ \citenamefont {Lein}(2014)}]{Barth2014}%
  \BibitemOpen
  \bibfield  {author} {\bibinfo {author} {\bibfnamefont {I.}~\bibnamefont
  {Barth}}\ and\ \bibinfo {author} {\bibfnamefont {M.}~\bibnamefont {Lein}},\
  }\href {\doibase 10.1088/0953-4075/47/20/204016} {\bibfield  {journal}
  {\bibinfo  {journal} {Journal of Physics B: Atomic, Molecular and Optical
  Physics}\ }\textbf {\bibinfo {volume} {47}},\ \bibinfo {pages} {204016}
  (\bibinfo {year} {2014})}\BibitemShut {NoStop}%
\bibitem [{\citenamefont {Moler}\ and\ \citenamefont {Loan}(1978)}]{Moler1978}%
  \BibitemOpen
  \bibfield  {author} {\bibinfo {author} {\bibfnamefont {C.}~\bibnamefont
  {Moler}}\ and\ \bibinfo {author} {\bibfnamefont {C.~V.}\ \bibnamefont
  {Loan}},\ }\href {http://epubs.siam.org/doi/abs/10.1137/1020098} {\bibfield
  {journal} {\bibinfo  {journal} {SIAM review}\ }\textbf {\bibinfo {volume}
  {20}},\ \bibinfo {pages} {801} (\bibinfo {year} {1978})}\BibitemShut
  {NoStop}%
\bibitem [{\citenamefont {Chirilă}\ \emph {et~al.}(2010)\citenamefont
  {Chirilă}, \citenamefont {Dreissigacker}, \citenamefont {van~der Zwan},\
  and\ \citenamefont {Lein}}]{Chirila2010}%
  \BibitemOpen
  \bibfield  {author} {\bibinfo {author} {\bibfnamefont {C.~C.}\ \bibnamefont
  {Chirilă}}, \bibinfo {author} {\bibfnamefont {I.}~\bibnamefont
  {Dreissigacker}}, \bibinfo {author} {\bibfnamefont {E.~V.}\ \bibnamefont
  {van~der Zwan}}, \ and\ \bibinfo {author} {\bibfnamefont {M.}~\bibnamefont
  {Lein}},\ }\href {\doibase 10.1103/PhysRevA.81.033412} {\bibfield  {journal}
  {\bibinfo  {journal} {Physical Review A}\ }\textbf {\bibinfo {volume} {81}},\
  \bibinfo {pages} {033412} (\bibinfo {year} {2010})}\BibitemShut {NoStop}%
\bibitem [{\citenamefont {Fano}(1985)}]{Fano1985}%
  \BibitemOpen
  \bibfield  {author} {\bibinfo {author} {\bibfnamefont {U.}~\bibnamefont
  {Fano}},\ }\href {\doibase 10.1103/PhysRevA.32.617} {\bibfield  {journal}
  {\bibinfo  {journal} {Physical Review A}\ }\textbf {\bibinfo {volume} {32}},\
  \bibinfo {pages} {617} (\bibinfo {year} {1985})}\BibitemShut {NoStop}%
\bibitem [{\citenamefont {Rza̧żewski}\ and\ \citenamefont
  {Piraux}(1993)}]{Rzazewski1993}%
  \BibitemOpen
  \bibfield  {author} {\bibinfo {author} {\bibfnamefont {K.}~\bibnamefont
  {Rza̧żewski}}\ and\ \bibinfo {author} {\bibfnamefont {B.}~\bibnamefont
  {Piraux}},\ }\href
  {http://journals.aps.org/pra/abstract/10.1103/PhysRevA.47.R1612} {\bibfield
  {journal} {\bibinfo  {journal} {Physical Review A}\ }\textbf {\bibinfo
  {volume} {47}},\ \bibinfo {pages} {1612} (\bibinfo {year}
  {1993})}\BibitemShut {NoStop}%
\bibitem [{\citenamefont {Zakrzewski}\ \emph {et~al.}(1993)\citenamefont
  {Zakrzewski}, \citenamefont {Delande}, \citenamefont {Gay},\ and\
  \citenamefont {Rza̧żewski}}]{Zakrzewski1993}%
  \BibitemOpen
  \bibfield  {author} {\bibinfo {author} {\bibfnamefont {J.}~\bibnamefont
  {Zakrzewski}}, \bibinfo {author} {\bibfnamefont {D.}~\bibnamefont {Delande}},
  \bibinfo {author} {\bibfnamefont {J.}~\bibnamefont {Gay}}, \ and\ \bibinfo
  {author} {\bibfnamefont {K.}~\bibnamefont {Rza̧żewski}},\ }\href
  {http://journals.aps.org/pra/abstract/10.1103/PhysRevA.47.R2468} {\bibfield
  {journal} {\bibinfo  {journal} {Physical Review A}\ }\textbf {\bibinfo
  {volume} {47}},\ \bibinfo {pages} {2468} (\bibinfo {year}
  {1993})}\BibitemShut {NoStop}%
\end{thebibliography}%

\end{document}